\documentclass[conference]{IEEEtran}
\IEEEoverridecommandlockouts

\usepackage{cite}
\usepackage{amsmath,amssymb,amsfonts}
\usepackage{algorithmic}
\usepackage{graphicx}
\usepackage{textcomp}
\usepackage{xcolor}
\usepackage{booktabs}
\usepackage{multirow}
\usepackage{url}
\usepackage{comment}
\usepackage{hyperref}

\def\BibTeX{{\rm B\kern-.05em{\sc i\kern-.025em b}\kern-.08em
    T\kern-.1667em\lower.7ex\hbox{E}\kern-.125emX}}

\begin{document}

\title{GitReq: A Gold Standard Dataset for Software Quality Requirements}
\author{\IEEEauthorblockN{Anonymous Authors}}

\author{
\IEEEauthorblockN{Farha Kamal}
\IEEEauthorblockA{\textit{Dept. of Computer Science} \\
\textit{Lamar University}\\
Beaumont, TX, USA \\
fkamal@lamar.edu}
\and
\IEEEauthorblockN{Md Humaun Kabir}
\IEEEauthorblockA{\textit{Dept. of Electrical and Computer Engineering} \\
\textit{Lamar University}\\
Beaumont, TX, USA \\
mkabir13@lamar.edu}
\and
\IEEEauthorblockN{Md Rakibul Islam}
\IEEEauthorblockA{\textit{Dept. of Computer Science} \\
\textit{Lamar University}\\
Beaumont, TX, USA \\
mislam108@lamar.edu}
}

\maketitle

\vspace{-0.5cm}
\begin{abstract}
GitHub issue trackers contain millions of developer-written quality concerns, including performance bottlenecks and security vulnerabilities, yet no publicly available GitHub dataset classifies these into fine-grained software quality categories. We construct and release GitReq (\textbf{Git}Hub \textbf{Req}uirement Issue), comprising 6,302 expert-validated requirements mined from 55,588 raw GitHub candidates across 4,080 repositories, labeled across eight ISO/IEC 25010:2011-aligned categories: Performance, Security, Portability, Availability, Fault-tolerance, Scalability, Maintainability, and a Functional baseline. Dataset construction involved category-specific triple-signal GitHub mining, separate non-functional requirement (NFR) and functional requirement (FR) preprocessing pipelines with per-category parameters, and expert human annotation achieving substantial inter-annotator agreement (Fleiss' Kappa~=~0.72). Zero-shot evaluation with four large language models (LLMs) establishes baselines, with GPT-5.2 reaching the highest macro-averaged F1 of 0.641. GitReq is publicly released with full  materials to advance research in automated requirement classification and software quality analysis.
\end{abstract}

\vspace{-0.3cm}
\section{Introduction}
\label{sec:introduction}

Software quality characteristics---performance, security, maintainability, and others---represent critical non-functional constraints in standards like ISO/IEC 25010:2011~\cite{iso25010,chung1999nfr,glinz2007nonfunctional}. Defects related to quality properties contribute 40--60\% of all software defects~\cite{boehm2001defects}, yet systematic classification of quality concerns in developer communications remains an open problem. Automated classification depends critically on labeled datasets, and recent advances in deep learning have demonstrated strong results on requirement classification tasks~\cite{hey2020norbert,luo2022prcbert}---but progress is constrained by four persistent limitations of existing datasets.

First, \textbf{scale}: existing labeled datasets are small, with the most widely used corpus containing only 625 requirements~\cite{cleland2007nfrdataset}, constraining model training and evaluation. Second, \textbf{domain diversity}: formal specification corpora draw requirements from narrow controlled settings such as student projects and curated Software Requirements Specification (SRS) documents~\cite{cleland2007nfrdataset,idate2024frnfr,islam2025decoding}, limiting generalizability to the broader open-source development landscape. Third, \textbf{incomplete coverage}: most datasets address either functional or non-functional categories, but not both within a single expert-validated corpus, preventing unified FR/NFR classification benchmarks. Fourth, and most critically, a \textbf{platform gap}: while large GitHub issue datasets exist for \textit{issue management intent} classification (bug/feature/question)~\cite{kallis2023nlbse}, no publicly available GitHub dataset provides expert-validated classification across multiple fine-grained software quality categories. GitHub issue trackers---where developers actively discuss, negotiate, and report quality concerns during development---are entirely absent from the requirements classification benchmark landscape.
We define a \textit{quality-bearing requirement} as any developer-authored GitHub issue expressing a quality constraint on system operation (non-functional) or a desired system behavior (functional), as judged by expert annotators against ISO/IEC 25010:2011 definitions~\cite{iso25010}. This operationalization is deliberate: developers rarely write formal ``shall'' statements in issue trackers---non-functional concerns appear as terse, implementation-driven phrases rather than structured specifications, while functional requirements require formal specification language (\texttt{shall}/\texttt{must}/user-story patterns) to be reliably distinguished from general feature discussion. This treatment of issue tracker content as requirements artifacts is consistent with prior work: Luo et al.~\cite{luo2022prcbert} mine Stack Overflow posts as NFR instances, establishing precedent for treating informal developer-authored text as a valid requirements source. This fundamental difference in how NFR and FR concerns surface in GitHub issues motivates the separate preprocessing pipelines described in Section~\ref{sec:preprocessing}.

\textbf{Our Contribution.} We present GitReq, a dataset of 6,302 GitHub issues expert-validated across eight ISO/IEC 25010:2011 quality categories~\cite{iso25010}, mined from 4,080 repositories spanning web frameworks, machine learning (ML) libraries, database systems, mobile applications, and cloud infrastructure. GitReq is constructed through a multi-stage pipeline: automated mining using category labels and keyword signals, separate NFR and FR preprocessing pipelines, and human expert validation by seven annotators (Fleiss' $\kappa = 0.72$). Of the 5,771 NFR requirements, 79.3\% were retrieved with a matching GitHub label (\texttt{found\_by\_label}~=~yes), confirming that weak supervision by developer-applied labels---not keyword heuristics alone---drives the NFR collection signal. The 531 functional requirements were filtered to 100\% formal specification language compliance by the FR preprocessing pipeline. Unlike GitHub issue corpora that classify issue management intent~\cite{kallis2023nlbse}, GitReq provides fine-grained quality classification. Unlike formal specification corpora~\cite{cleland2007nfrdataset,ferrari2017detecting}, GitReq reflects authentic developer expression in implementation-driven issue discussions. We release the dataset at \url{https://doi.org/10.6084/m9.figshare.31669477}.
\vspace{-0.2cm}
\section{Dataset Construction}
\label{sec:methodology}
\begin{figure*}[!t]
  \centering
  \includegraphics[width=\textwidth, trim=0 22.5cm 0 0.5cm, clip]{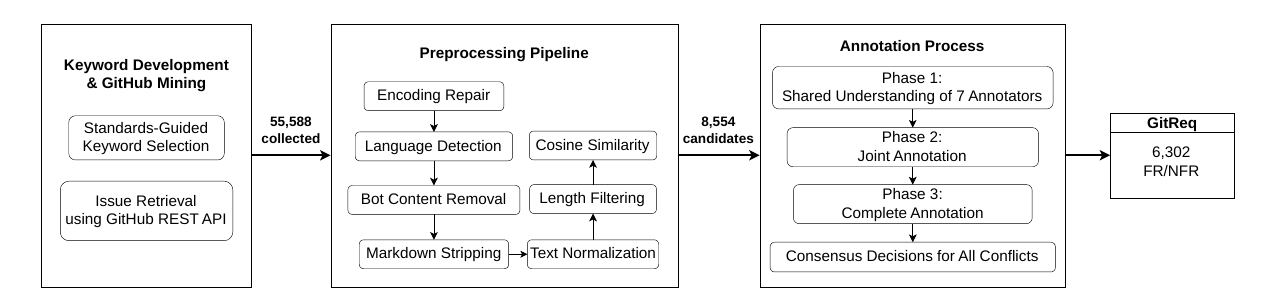}
  \caption{GitReq construction workflow: category-specific mining $\rightarrow$ NFR/FR preprocessing pipelines $\rightarrow$ expert annotation $\rightarrow$ 6,302 validated requirements.}
  \vspace{-0.4cm} 
  \label{fig:workflow}

\end{figure*}

\vspace{-0.2cm}
GitReq was built through four stages: (1)~category-specific GitHub mining, (2)~separate NFR and FR preprocessing pipelines, (3)~expert human annotation, and (4)~consensus-based quality filtering. Figure~\ref{fig:workflow} illustrates the pipeline.

\vspace{-0.3cm}
\subsection{Requirement Categories}
\vspace{-0.2cm}
GitReq covers seven NFR categories and one Functional category. NFR categories follow ISO/IEC 25010:2011~\cite{iso25010}: \textit{Performance}, \textit{Security}, \textit{Portability}, \textit{Availability}, \textit{Fault-tolerance}, \textit{Maintainability}, and \textit{Scalability}. Although ISO/IEC 25010:2011 does not list Scalability as a separate top-level characteristic, we treat it as distinct from Performance in practice: Performance concerns response time and resource behavior under a given workload, whereas Scalability concerns the system's ability to sustain acceptable behavior as workload or concurrency increases. This distinction was reflected in keyword design and validated during annotation.
\vspace{-0.2cm}
\subsection{Keyword Development}
\vspace{-0.2cm}
Category-specific keywords were developed using three sources: (1)~established requirements engineering literature~\cite{chung1999nfr,glinz2007nonfunctional}, (2)~manual inspection of GitHub issues to identify developer phrasing, and (3)~iterative pilot mining. NFR keywords span both technical terminology (e.g., \texttt{latency}, \texttt{encryption}, \texttt{failover}) and natural language developer expressions (e.g., \texttt{runs slowly}, \texttt{data leak}). Final keyword counts ranged from 10 (Availability) to 55 
(Portability) across NFR categories; Functional requirements used 87 
keywords targeting system behaviors and user actions (e.g., 
\texttt{feature}, \texttt{allow user}, \texttt{system should}). Complete keyword lists are provided in the materials package~\cite{gitreq2026}.
\subsection{GitHub Mining}
\label{sec:mining}
\vspace{-0.2cm}
We mined public GitHub issues via the GitHub REST API~\cite{github2024api}, using category-adaptive strategies based on label availability. Bot-authored issues were excluded during retrieval; duplicates from repeated keyword matches were removed by deduplicating on (repository, issue number, keyword, generic label). Bug reports were excluded at query time by requiring all issues to carry a requirement-intent label (\texttt{feature request}, \texttt{enhancement}, or \texttt{user story}); bug-typed issues do not carry these labels and therefore do not enter the candidate pool. Across all categories and strategies, initial collection yielded 55,588 candidate issues. Table~\ref{tab:mining} summarizes raw and final counts per category where \textit{Raw}: total issues retrieved before preprocessing; \textit{Final}: issues retained after preprocessing and expert annotation.

\begin{table}[t]
\centering
\caption{Mining Strategy and Issue Counts per Category.}
\label{tab:mining}
\small
\begin{tabular}{lrrl}
\toprule
\textbf{Category} & \textbf{Raw} & \textbf{Final} & \textbf{Mining Strategy} \\
\midrule
Security        &  8,352 & 1,646 & Triple-signal (standard) \\
Performance     &  4,666 & 1,509 & Triple-signal (standard) \\
Portability     &  3,754 & 1,284 & Triple-signal (standard) \\
Availability    & 12,281 &   738 & Multi-strategy (3 stages) \\
Fault-tolerance &  1,061 &   293 & Multi-label + multi-extract \\
Scalability     &    176 &   157 & Label + multi-extract \\
Maintainability &    264 &   144 & Label + multi-extract \\
Functional      & 25,034 &   531 & Generic labels only \\
\midrule
\textbf{Total}  & \textbf{55,588} & \textbf{6,302} & \\
\bottomrule
\end{tabular}
\vspace{-0.6cm}
\end{table}

\subsubsection{Standard Triple-Signal Mining (Security, Performance, Portability)}
For categories with well-established GitHub labels, retrieval combined three signals: (1)~a category-specific GitHub label (e.g., \texttt{label:security}), (2)~one of three generic requirement-intent labels (\texttt{"feature request"}, \texttt{"enhancement"}, \texttt{"user story"}), and (3)~a category keyword in the issue title or body. A representative query takes the form:
\begin{center}
\texttt{label:security label:"feature request" type:issue} 
\texttt{"authentication" in:title,body}
\end{center}
Security used 43 keywords yielding 8,352 raw issues; Performance used 41 keywords yielding 4,666; Portability used 55 keywords yielding 3,754. For Security, Performance, and Fault-tolerance, 100\% of final requirements were retrieved with a matching category label on the issue (\texttt{found\_by\_label}~=~yes). For Portability, a set of seven category-specific labels (\texttt{portability}, \texttt{cross-platform}, \texttt{compatibility}, \texttt{platform-support}, \texttt{os-support}, \texttt{docker}, \texttt{containerization}) was used; 830 of 1,284 final Portability requirements (64.6\%) carried a portability-specific label, while the remaining 454 (35.4\%) were retrieved via generic label + portability keyword, reflecting common cross-platform issues filed without a dedicated portability label. Across all seven NFR categories combined, 79.3\% of final requirements were confirmed by a matching category label, with the remainder relying on keyword and related-label signals.

\subsubsection{Multi-Label Mining (Fault-tolerance)}
Because GitHub repositories use varied labels for fault-tolerance concepts, retrieval used four related category labels (\texttt{fault-tolerance}, \texttt{error-handling}, \texttt{resilience}, \texttt{reliability}) combined with generic labels and 40 keywords, yielding 1,061 raw issues.

\subsubsection{Multi-Strategy Mining (Availability)}
The \texttt{availability} label is rare on GitHub; initial retrieval using category label + generic labels + keywords returned only 7 issues. We therefore deployed three complementary strategies: (1)~related labels (\texttt{reliability}, \texttt{infrastructure}, \texttt{ops}, \texttt{devops}, \texttt{deployment}) with availability keywords; (2)~generic requirement-intent labels with keywords but no category label; and (3)~keyword-only searches with no label requirement. This yielded 12,281 raw candidates. In the final 738 Availability requirements, the generic-label strategy contributed 79.1\%, with related-label and keyword-only 
strategies accounting for the remainder---confirming that Availability is the one category where keyword and related-label signals dominate rather than category-label confirmation.
\subsubsection{Small-Category Mining (Maintainability, Scalability)}
Both \texttt{maintainability} and \texttt{scalability} labels are rare on GitHub, yielding only 264 and 176 raw issues respectively through standard triple-signal retrieval.

\subsubsection{Functional Requirements Mining}
FR requirement candidates were mined using only the three generic requirement-intent labels with 87 functional keywords, with no category label required. These labels serve solely as a collection filter to identify issues expressing desired system behaviors; the actual FR classification is determined entirely by the formal-language scoring rubric and expert annotation, not by the mining label. To reduce NFR contamination, issues containing quality-oriented terminology (e.g., \texttt{performance}, \texttt{security}, \texttt{scalability}, \texttt{availability}, \texttt{reliability}) were excluded at query time. This yielded 25,034 raw candidates.

\subsection{Preprocessing Pipelines}
\label{sec:preprocessing}

Raw GitHub issues contain substantial noise: templates, code blocks, URLs, markdown artifacts, and conversational discussion. We designed \textit{separate} preprocessing pipelines for NFR and FR categories, reflecting a fundamental difference in how these two requirement types surface in GitHub issues. NFR concerns appear as terse, implementation-driven expressions that rarely follow formal grammar (e.g., ``uptime monitoring ping url check for 200''), so the NFR pipeline focuses on \textit{semantic content}---category keywords, requirement-indicating verbs, and appropriate length---without requiring grammatical sentence structure. Functional requirements, by contrast, must be distinguished from the vastly larger population of feature suggestions, questions, and boilerplate; without a formal specification signal (\texttt{shall}/\texttt{must}/user story), no reliable separation is possible. The FR pipeline therefore enforces formal language compliance as a prerequisite for inclusion.

\subsubsection{Common Preprocessing Steps}
Both pipelines began with: encoding repair using \texttt{ftfy}~\cite{ftfy}, markdown and HyperText Markup Language (HTML) removal, fenced and inline code block stripping, whitespace normalization, and English language filtering via \texttt{langdetect}~\cite{langdetect}. Bot-authored issues were excluded via username pattern matching.

\subsubsection{NFR Preprocessing}
Each category applied per-category length windows and Term Frequency–Inverse Document Frequency (TF-IDF) cosine similarity deduplication at a threshold of 0.85, which reliably identifies substantive textual overlap while preserving distinct issues, consistent with deduplication practices in NLP pipelines~\cite{niarchos2022semantic}. 
Varying the threshold between 0.80 and 0.90 produced less than 3\% difference in retained candidates, confirming stability of the chosen value. Crucially, all heuristic filtering served only to reduce the candidate pool for human review; no text entered the final dataset without independent expert validation, limiting the risk of systematic bias from any single preprocessing rule.
Standard categories (Security, Performance, Portability) used single-pass sentence extraction with output lengths of 30--100 words. Availability required three-stage filtering---smart bug removal, strict preprocessing, and ultra-strict language filtering---reducing 12,281 candidates to 738 (94\% reduction). Sparse categories (Fault-tolerance, Maintainability, Scalability) combined single-extraction and multi-extraction passes, where structured list items within issues were segmented into individual requirement instances to maximize recall; similarity deduplication was applied after each pass before annotation.

\subsubsection{FR Preprocessing Pipeline}
The FR pipeline required fundamentally different criteria. Issue \textit{bodies} were extracted separately from titles (titles typically contain only metadata like ``[Feature Request]''). Candidate sentences were scored based on the presence of formal specification markers---\texttt{shall}, \texttt{must}, user story format (``As a [role], I want to\ldots''), and user-capability expressions (\texttt{users can}/\texttt{user can})---with action verbs (\texttt{allow}, \texttt{display}, \texttt{store}, \texttt{validate}) contributing incrementally. Sentences with questions, hedging, informal language, templates, past tense, bug language, or NFR terminology received penalties. A minimum score threshold was required for passage. This first-pass preprocessing used output length 15--70 words with deduplication at 0.85, following the same threshold justification~\cite{niarchos2022semantic}.

A second-pass filtering applied 14 independent rejection rules covering: questions (\texttt{?}, \texttt{should we}); informal or proposal language (\texttt{I think}, \texttt{I would}, \texttt{we need}); hedging (\texttt{maybe}, \texttt{might}, \texttt{would be nice}); template and markdown remnants (\texttt{Feature:}, \texttt{**Header**}); current-state or past-tense language (\texttt{Currently}, \texttt{was}, \texttt{were}); social text (\texttt{thanks}, \texttt{sorry}); bug language (\texttt{bug}, \texttt{error}); NFR terminology (\texttt{performance}, \texttt{security}); missing sentence structure; and length outside 15--60 words. Passing required at least one formal specification marker: \texttt{shall}, \texttt{must}, \texttt{will}, user story pattern, or \texttt{users can}. This two-stage pipeline reduced 25,034 raw candidates to 620 validated functional requirements entering annotation phase.\\ After all preprocessing steps, 8,554 candidate texts remained and were forwarded to human annotation. Detailed per-stage preprocessing counts for all categories are provided in the materials package~\cite{gitreq2026}.

\begin{table*}[t]
\centering
\caption{Category Distribution, Text Statistics, and Representative Examples}
\label{tab:stats}
\footnotesize
\begin{tabular}{lrrrr p{10cm}}
\toprule
\textbf{Category} & \textbf{N} & \textbf{\%} & \textbf{Avg Wds} & \textbf{Avg Chrs} & \textbf{Examples} \\
\midrule
Security        & 1,646 & 26.1 & 61.4 & 416.9 & Add encryption for SAML 2.0 tokens for WS-Fed clients; OAuth configuration for SAP BTP Cloud Foundry deployments \\
Performance     & 1,509 & 23.9 & 63.9 & 410.8 & Investigate slowness of DBS3Upload component under heavy load; Reduce initial lag after loading the map \\
Portability     & 1,284 & 20.4 & 49.4 & 317.1 & Add support for Alpine 3.14 for .NET Core 2.1, 3.1 and .NET 5; Add support for Ubuntu 25.04 for .NET 8, 9, and 10 \\
Availability    &   738 & 11.7 & 41.7 & 275.9 & Parallelized zero-downtime restart to avoid load spikes; Non-blocking health check for load balancers \\
Fault-tolerance &   293 &  4.6 & 40.3 & 293.3 & Implement circuit breaker pattern for failing feeds; Add fallback to in-memory limiter if Redis is unavailable \\
Scalability     &   157 &  2.5 & 34.7 & 240.9 & Distributed rate limiting with Redis to support multi-instance deployments; Implement sampling strategies, sampling metadata in records \\
Maintainability &   144 &  2.3 & 42.6 & 282.3 & Refactor config objects with explicit attributes and defaults; Rename and organize .py files following a consistent system \\
Functional      &   531 &  8.4 & 29.5 & 177.0 & The system must provide option to export notes as separate file; As a user, I want to cycle through images in a listing; the order must not change \\
\midrule
\textbf{Total}  & \textbf{6,302} & \textbf{100} & \textbf{52.5} & \textbf{345.2} & \\
\bottomrule
\end{tabular}
\vspace{-0.3cm}
\end{table*}

\subsection{Human Annotation and Validation}
\subsubsection{Annotator Description}
Seven expert annotators participated in the validation task, selected for domain expertise in software engineering and requirements engineering. Among them, one holds PhD in Software Engineering with six years of research experience in software engineering. The remaining six possessed 3--5 years of professional software engineering experience with demonstrated expertise in requirements elicitation (gathering stakeholder needs), software quality assurance, and testing. All annotators were affiliated with research universities in North America.

\subsubsection{Structured Training}
We conducted structured training across three phases~\cite{islam2024fourdimension}.

\textbf{Phase 1} (shared understanding) provided comprehensive instruction on each category, including ISO/IEC 25010:2011 definitions and ten representative examples per category drawn from a pilot annotation of 100 texts selected via stratified random sampling (12--13 per category, balanced regardless of mining-stage proportions), excluded from the final dataset. This three-hour session ensured clarity before annotation commenced.

\textbf{Phase 2} (joint annotation) had all seven annotators jointly annotate 200 identical texts under supervision, enabling real-time discussion of ambiguous cases and establishing consistent application of definitions. Annotators then independently labeled 200 unique texts; Fleiss' Kappa~\cite{fleiss1971measuring} ranged 0.62--0.75 across categories, confirming substantial agreement. Following successful training, annotators proceeded to the primary corpus of 8,554 texts.

\subsubsection{Completing the Annotation (Phase 3)}
The remaining texts were distributed between two teams of three and four members; each team received approximately half the texts. Within each team, every member independently annotated all assigned texts (three or four independent labels per text), with no intra-team discussion permitted. Upon completion, Fleiss' Kappa~\cite{fleiss1971measuring} ranged from 0.62 (Scalability) to 0.75 (Performance), with overall Kappa of 0.72. Majority classes achieved higher agreement (Performance: 0.75, Security: 0.74, Functional: 0.70) than minority classes (Scalability: 0.62, Maintainability: 0.64), consistent with prior annotation studies~\cite{ferrari2017detecting,islam2024fourdimension}. Labels were determined through majority voting. When no majority existed (e.g., 2--2 splits in Team 2, or complete disagreement), these cases were flagged for further discussion.

\subsubsection{Resolving Disagreements}
This review identified 2,437 texts requiring reconsideration (no majority consensus). Through collective discussion, 285 achieved consensus and were retained; the rest were excluded due to: multiple distinct quality concerns incompatible with single-label classification (n=706), insufficient clarity for reliable classification (n=803), primarily code content (n=421), or unresolved disagreement (n=222).
The final quality-filtered dataset comprises 6,302 expert-validated classified texts, yielding 74.5\% retention from the annotation phase and 73.7\% retention from preprocessed candidates (8,554 preprocessed $-$ 100 pilot exclusions $-$ 2,152 annotation exclusions = 6,302).
\vspace{-0.2cm}
\section{Dataset Characteristics}
\label{sec:dataset}

\begin{table*}[t]
\centering
\caption{Zero-Shot LLM Results on GitReq (P = Precision, R = Recall, F1 = F1-Score)}
\label{tab:results}
\footnotesize
\begin{tabular}{lcccccccccccc}
\toprule
& \multicolumn{3}{c}{\textbf{LLaMA 3 70B}} & \multicolumn{3}{c}{\textbf{Gemma 3 4B}} & \multicolumn{3}{c}{\textbf{Mistral 7B}} & \multicolumn{3}{c}{\textbf{GPT-5.2}} \\
\cmidrule(lr){2-4}\cmidrule(lr){5-7}\cmidrule(lr){8-10}\cmidrule(lr){11-13}
\textbf{Class} & P & R & F1 & P & R & F1 & P & R & F1 & P & R & F1 \\
\midrule
Security        & 0.945 & 0.800 & 0.867 & 0.939 & 0.526 & 0.674 & 0.961 & 0.476 & 0.637 & 0.940 & 0.880 & 0.909 \\
Performance     & 0.950 & 0.808 & 0.873 & 0.879 & 0.694 & 0.776 & 0.940 & 0.427 & 0.587 & 0.960 & 0.883 & 0.920 \\
Portability     & 0.964 & 0.463 & 0.626 & 0.980 & 0.075 & 0.139 & 0.973 & 0.362 & 0.527 & 0.943 & 0.657 & 0.774 \\
Availability    & 0.931 & 0.535 & 0.680 & 0.981 & 0.142 & 0.249 & 0.992 & 0.179 & 0.303 & 0.977 & 0.458 & 0.624 \\
Fault-tolerance & 0.714 & 0.702 & 0.708 & 0.428 & 0.212 & 0.283 & 0.556 & 0.307 & 0.396 & 0.599 & 0.590 & 0.595 \\
Scalability     & 0.423 & 0.382 & 0.401 & 0.484 & 0.293 & 0.365 & 0.228 & 0.439 & 0.300 & 0.476 & 0.446 & 0.461 \\
Maintainability & 0.413 & 0.694 & 0.518 & 0.080 & 0.799 & 0.145 & 0.518 & 0.507 & 0.512 & 0.196 & 0.903 & 0.322 \\
Functional      & 0.243 & 0.876 & 0.381 & 0.183 & 0.791 & 0.297 & 0.143 & 0.962 & 0.248 & 0.396 & 0.770 & 0.523 \\
\midrule
\textbf{Macro F1} & \multicolumn{3}{c}{0.632} & \multicolumn{3}{c}{0.366} & \multicolumn{3}{c}{0.439} & \multicolumn{3}{c}{\textbf{0.641}} \\
\bottomrule
\end{tabular}
\vspace{-0.4cm}
\end{table*}
Table~\ref{tab:stats} presents the full distribution and text statistics. Security (26.1\%) and Performance (23.9\%) together comprise 50\% of the dataset, reflecting practitioners' focus on threat protection and system responsiveness in open-source development. Portability (20.4\%) reflects the cross-platform demands inherent to open-source software. Availability (11.7\%) and Fault-tolerance (4.6\%) represent reliability concerns. Maintainability (2.3\%) and Scalability (2.5\%) appear less frequently, likely emerging later in project lifecycles or addressed architecturally. Functional requirements (8.4\%) serve as a comparative baseline.

Security and Performance requirements average over 60 words, consistent with the 40--100-word preprocessing filter. Functional requirements are most concise (29.5 words), reflecting strict FR pipeline criteria. All 531 functional requirements contain at least one formal marker (\texttt{must} 78.9\%, \texttt{shall} 10.2\%, \texttt{will} 7.7\%, user-story 14.3\%, \texttt{users can} 12.1\%; percentages sum to $>$100\% as requirements may contain multiple markers). The vocabulary spans 15,274 unique words, reflecting the linguistic diversity of open-source developer discourse.

GitReq spans 4,080 unique repositories (3,644 NFR, 455 FR, with 19 overlap). The median repository contributed a single requirement text, and 97.5\% contributed five or fewer, ensuring cross-project breadth rather than concentration in a few large repositories.

Table~\ref{tab:stats} includes a representative example per category. The range from abbreviated informal NFR phrasing to formal FR specification language distinguishes GitReq from both raw issue corpora and formal SRS datasets such as PROMISE and PURE.

\vspace{-0.2cm}
\section{Zero-Shot LLM Evaluation: A Use Case}

\label{sec:evaluation}
To illustrate GitReq's utility as a benchmark, we evaluate four LLMs under a zero-shot classification setting: GPT-5.2 (model string: \texttt{gpt-5.2})~\cite{openai2025gpt5}, LLaMA~3~70B~\cite{touvron2023llama}, Mistral~7B~\cite{jiang2023mistral}, and Gemma~3~4B~\cite{google2024gemma}. Each model was prompted to assign one of the eight ISO/IEC 25010:2011-aligned quality categories to each requirement text, using category definitions drawn directly from the standard and no labeled examples (temperature~=~0). The prompt template used for evaluation is included in the dataset package~\cite{gitreq2026}. 
Our zero-shot evaluation protocol follows the methodology established by Alhoshan et al.~\cite{alhoshan2023zeroshot}, who applied zero-shot learning to requirements classification across three datasets, demonstrating that zero-shot evaluation provides a training-free assessment of dataset difficulty and model capability without requiring task-specific fine-tuning or training splits. Following this precedent, we evaluate on the full GitReq corpus of 6,302 texts under the same zero-shot paradigm, which introduces no data leakage.
Table~\ref{tab:results} reports per-class precision, recall, and F1 alongside macro F1. GPT-5.2 achieves the highest macro F1 (0.641), with strong results on Performance and Security across all models. Notably, Maintainability F1 varies widely across models (0.145--0.518): Gemma achieves high recall (0.799) but very low precision (0.080), suggesting substantial over-prediction of this category, while GPT-5.2 shows the inverse pattern (precision 0.196, recall 0.903). Sparse categories (Maintainability, Scalability) and high-noise pools (Availability, Functional) remain challenging, confirming GitReq as a non-trivial benchmark for future classification research.
\vspace{-0.2cm}
\section{Research Opportunities}
\vspace{-0.2cm}
GitReq addresses the platform gap by providing the first expert-validated GitHub dataset for fine-grained quality classification, enabling research directions that SRS-based corpora could not support. Researchers can train and benchmark supervised classifiers, including parameter-efficient fine-tuned transformers on authentic developer language rather than formal specification text, producing models that generalize to real issue tracking workflows. The dataset supports cross-platform transfer learning studies comparing model behavior across formal corpora (PROMISE, PURE), community-platform collections (NFR-SO), and operational issue trackers, directly probing how writing style affects classification performance~\cite{kamal2025robust}. GitReq also enables development of automated quality coverage tools that flag under-addressed quality concerns during active development, integrating into CI/CD pipelines or issue triage workflows. Finally, the zero-shot baselines established here invite deeper LLM analysis: prompt sensitivity, few-shot adaptation, and error pattern studies across the eight quality categories.

\vspace{-0.4cm}
\section{Related Datasets}
\vspace{-0.2cm}
Requirement datasets divide into formal specification corpora, community-platform collections, and label-extension efforts. GitReq differs from all three in both text source and annotation method.

\textbf{Formal Specification Corpora.} The PROMISE NFR dataset~\cite{cleland2007nfrdataset} contains 625 labeled requirements (255 FR, 370 NFR) from student software projects, released via the RE'17 Data Challenge, and has served as the primary benchmark for over a decade. Lima et al.~\cite{lima2019promise} extended it with PROMISE\_exp (969 instances, 444 FR and 525 NFR) by adding requirements from 34 additional SRS documents. Ferrari et al.~\cite{ferrari2017detecting} released PURE, a collection of 79 publicly available requirements \textit{documents} (34,268 sentences); PURE is a raw corpus rather than a labeled classification dataset. Idate and Rao~\cite{idate2024frnfr} drew on PURE and additional SRS documents to produce FR\_NFR, a labeled dataset of 6,118 requirements (3,964 FR, 2,154 NFR) for binary classification. All share the same limitation: text comes from formal SRS documents where ``shall/must'' language is the norm, not from operational developer workflows.

\textbf{Community-Platform Collections.} Luo et al.~\cite{luo2022prcbert} introduced NFR-SO, mined from Stack Overflow across seven NFR categories, used to train and evaluate PRCBERT. Baker et al.~\cite{baker2019automatic} evaluated artificial neural network (ANN) and convolutional neural network (CNN) classifiers on $\sim$914 instances combining PROMISE and IEEE RE 2017 data across five NFR categories. The limitation of community-mined collections is label validity: Stack Overflow tags (e.g., \textit{performance}, \textit{security}) reflect \textit{topic interest}, not \textit{requirement intent}---a post tagged \textit{performance} may be a debugging question rather than a quality constraint. GitReq addresses this by requiring each issue to carry both a quality category label \textit{and} a requirement-intent label before annotation.

\textbf{Label-Extension Efforts.} Rejith Kumar et al.~\cite{rejithkumar2025nice} released NICE (ICSE 2025), re-annotating augmented PROMISE with multi-label annotations across 12 NFR categories. NICE advances annotation expressiveness but inherits the SRS text source of PROMISE. No existing dataset combines GitHub issue provenance, expert validation, and fine-grained FR/NFR quality classification. GitReq addresses this gap.
\section{Limitations}
GitReq is limited to English-language GitHub issues, which may not represent multilingual or industrial contexts. Some subjectivity remains for overlapping categories: Scalability (n~=~157, $\kappa$~=~0.62) has the lowest agreement and smallest size; researchers may consider merging it with Performance for studies requiring high-confidence labels. 
Availability (n~=~738), Fault-tolerance, and Maintainability relied on keyword-based or multi-strategy retrieval due to sparse GitHub label coverage; however, keywords determined only the candidate pool---expert annotators independently excluded issues that did not clearly express a quality constraint. The imbalanced distribution---particularly Scalability (n~=~157) and Maintainability (n~=~144)---may require stratified sampling, class weighting, or oversampling strategies in supervised settings.
\section{Conclusion}
We presented \textbf{GitReq}, a large-scale expert-validated dataset of 6,302 GitHub issues annotated across eight quality categories, constructed from 55,588 raw candidates across 4,080 repositories through category-specific triple-signal mining, separate NFR and FR preprocessing pipelines with per-category parameters, and human expert validation (Fleiss' Kappa~=~0.72). Of the 5,771 NFR requirements, 79.3\% were confirmed by a matching category GitHub label; the 531 functional requirements achieved 100\% formal specification language compliance. Zero-shot evaluation with four LLMs establishes baselines (macro F1~=~0.641), confirming GitReq as a non-trivial benchmark. GitReq fills the gap between issue-type classification corpora and formal specification datasets, providing a platform-native quality classification resource for multilingual expansion, fine-tuned classification benchmarks, and integration with pull request or code review corpora~\cite{gitreq2026}.
\vspace{-0.4cm}

\end{document}